\documentclass[twocolumn]{aastex631}
\usepackage{amsmath,amsfonts,amssymb,graphicx,chngcntr,multirow, float,booktabs, afterpage}
\usepackage[]{hyperref}
\hypersetup{colorlinks=true}

\received{}
\revised{}
\accepted{}

\shorttitle{Black hole accretion disks with outflows}

\begin{document}

\title{  Black hole accretion disks with outflows \\ II. Time dependent Green's function solutions in Newtonian gravity }
\author{Andrew Mummery} 
\affiliation{School of Natural Sciences, Institute for Advanced Study, 1 Einstein Drive, Princeton, NJ 08540, USA}
\email{amummery@ias.edu}

\date{\today}

\label{firstpage}

\begin{abstract} 
We present Green's function solutions of the Newtonian time-dependent thin disk equations in the presence of outflows, showing that simple and exact analytical expressions exist in various natural limits of the problem. These Green's functions are mathematically very similar to the classical Lynden-Bell \& Pringle solutions in the absence of outflows, but differ strongly in their precise physical details and observational implications. Solutions are presented for phenomenological radius-dependent outflows which both do and do not torque the local accretion flow, and for outflows which are launched proportional to the local accretion rate. Generically, outflows lead to a more rapid decay of the bolometric luminosity of the disk, flatten the radial dependence of the disk temperature, and suppress variability in the accretion rate at small radii and low frequencies (on long timescales). Observational implications of these four results are discussed in detail. 
\end{abstract}

\noindent

\section{Introduction} 
The accretion of material onto black holes is very often associated with outflows, a statement for which there is a large body of observational evidence \citep[see e.g.,][for a review in the context of X-ray binaries, or \citealt{elvis_structure_2000} for an AGN discussion]{Darias26}. Outflows are also routinely seen in numerical simulations of the black hole accretion process \citep[see e.g.,][for two recent simulations in different regimes which both show profound outflows]{Zhang25, Guo25}. Despite this, analytical models (i.e., those routinely fit to data) generally neglect the impact of these outflows on the disk structure, a statement which applies both to steady-state solutions \citep[e.g.,][]{PageThorne74} as well as those black hole disks which evolve with time \cite[e.g.,][]{Mummery23a}. It is the purpose of this series of papers to perform a systematic theoretical analysis of the properties of thin accretion disks in the presence of outflows. In this paper, we examine the time evolution of thin accretion flows in a Newtonian gravity in the presence of outflows. 

In the absence of outflows, Green's function solutions of the Newtonian thin disk evolution equations go back to \cite{LBP74}, under the assumption of simple radius-dependent ``viscosity''  (in reality turbulent stress) profiles. A Green’s function is an extremely powerful analytical tool, as it describes the evolution of the disk resulting from an impulsive ($\delta$-function) initial condition. The solution of the general problem defined by an arbitrary initial condition or source of matter is then simply given by an integral of this initial condition with the Green’s function solution, opening up a wide range of physical problems to simple analysis. 

In \cite{Mummery23a} the generalization of these Newtonian Green's function solutions to the relativistic regime was made, where a derivation of a set of Green's function solutions valid in the full Kerr geometry (again for parameterized turbulent stresses) was presented. These relativistic Green's functions are built from combining various asymptotic limits, one of which being the Newtonian limit. As such, before Green's function solutions of the relativistic thin disk equations in the presence of outflows can be constructed, the equivalent problem in Newtonian gravity must first be solved. It is the purpose of this paper to present Green's function solutions of the Newtonian thin disk equations for as general a set of outflow profiles as possible. Future papers in this series will generalize these results to the Kerr geometry. 

In this paper we seek to be as general as possible in our parametrisations of the properties of the outflows from the thin accretion flow.  The reason we wish to consider as general a set of outflow profiles as is analytically tractable is for the following, simple, reason. Outflows can be launched due to a wide variety of different mechanisms, for each of which a different ``natural'' parameterisation may well present itself. These include winds launched by magnetohydrodynamic effects, an idea going back to \cite{Blandford82} \citep[see e.g.,][for a discussion of a possible parameterisation of MHD winds in the proto planetary disk literature, and some self-similar asymptotic solutions in \citealt{Tabone22}]{Suzuki16}, or they could be line driven  \citep[e.g.,][]{Laor14} and will therefore depend on the local disk flux at certain frequencies, or they could be driven by radiation pressure \citep[see e.g., the simulations of][]{Zhang25}, particularly in super-Eddington flows, where the most natural parameterisation may be in terms of the local disk accretion rate. Other routes to disk outflows include thermal pressure driving \cite[e.g.,][]{proga2007}, or even dust opacity effects \cite[really another example of line driving][]{czerny2011}. As far as is possible we attempt to avoid selecting any one driving mechanism in this paper. 

Disk winds can, and also may not, provide a torque onto the accretion flow, depending on whether fluid elements change their angular momentum over their ejection from the disk atmosphere. One zone models \citep[e.g.,][]{Metzger08} typically allow the torque on the flow from this wind to be a tuneable parameter, as do some global parameterizations \citep[e.g.,][]{Suzuki16, Tabone22}. It is likely that different wind driving mechanisms provide different ``natural'' torque profiles on the flow. Indeed, some recent simulations \citep[e.g.,][]{Guo25} suggest that often disk winds do not in fact torque the disk to an appreciable degree. We follow a similar course here, allowing winds to, as far as is tractable, have tuneable torque profiles. Exact Green's function solutions can be found in the presence or absence of torques.  

The layout of this paper is as follows. In section 2 we present a general preliminary analysis, presenting the form of the governing equations, and showing how one constructs a Green's function solution. In section 3 and 4 the Green's function solutions of the Newtonian disk equation in the presence of outflows are presented, for phenomenological radius-dependent outflow profiles which do (section 4), or do not (section 3) provide a torque on the disk. In section 5 we consider perhaps the most natural parameterisation of the outflow, in terms of the wind efficiency $\varepsilon_w$, defined by 
\begin{equation}
    \varepsilon_w \equiv - {\pi r^2 F_M(r, t) \over \dot M_{\rm acc}(r, t)}, 
\end{equation}
where $F_M$ is the local rate at which the disk surface density is removed from the flow and $\dot M_{\rm acc}(r, t)$ is the local accretion rate. Green's function solutions of the thin disk equations exist in the limit where $\varepsilon_w$ is a global constant. 

In section 6, we present the form of the Fourier-Green's functions, and discuss how outflows impact the variability of the accretion rate in thin disks. Observational implications of all of the properties of the solutions derived in this work are presented in section 7. We conclude in section 8. 

The following subsections are in many places rather technical, although they ultimately result in simple final results. The reader who is interested in observational implications of these results may wish to skip directly to section 7, where a detailed discussion of the impacts of disk outflows on observables is made. 

\section{General considerations}
\subsection{Newtonian limit and simplifying assumptions}
The general relativistic thin disk evolution equation in the presence of outflows (as derived in Paper I) takes the following general form 
\begin{multline}
    U^0 {\partial \Sigma \over \partial t} - {1\over r} {\partial \over \partial r}\left[{U^0 \over U_\phi '} {\partial \over \partial r} \left({r\Sigma W^r_\phi \over U^0}\right)\right] \\ = - F_M + {1\over r}{\partial \over \partial r}\left({r \over U_\phi'}F_J\right).
\end{multline}
Here notation is standard, $r$ and $t$ are the radial and temporal (Boyer-Lindquist) coordinates respectively, $\Sigma$ is the disk surface density, and $W^r_\phi$ is the disk turbulent stress, $U^0$ is the temporal component of the disk 4-velocity,  $U_\phi$ is the fluid's specific angular momentum  and a $'$ denotes a radial derivative. The left hand side of this equation is just the standard thin disk relativistic evolution equation \citep{Balbus17}, the right side contains two additions sourced from outflows. Namely the two fluxes out of the disk are the mass flux
\begin{equation}
    F_M \equiv \int{\partial \over \partial z}(\rho U^z)\, {\rm d}z, 
\end{equation}
and angular momentum flux
\begin{equation}
    F_J \equiv \int \rho U^z{\partial \over \partial z}(U_\phi)\, {\rm d}z .
\end{equation}
Here, $\rho$ is the disk density and $U^z$ the vertical (i.e., $z$ coordinate) velocity. 

The Newtonian limit corresponds to taking $U^0 = 1$ (the Lorentz factor of the flow) and $U_\phi = \sqrt{GMr}$ (the specific angular momentum of the flow). 

We will make as simplifying assumptions the following statements
\begin{equation}
    F_M \approx 2 \rho(H) U^z(H) \equiv 2 \Sigma/t_w, 
\end{equation}
where we have defined the ``wind timescale'' as 
\begin{equation}
    t_w \equiv H/U^z(H),
\end{equation}
and the first line assumes symmetry above and below the disk plane. As of yet this makes no assumptions about the physics of $t_w$, which we shall vary throughout this paper.

Similarly the angular momentum flux is assumed to take the form
\begin{equation}
    F_J \approx 2\rho(H) U^z(H) \left[U_\phi(H) - U_\phi(0)\right] \equiv 2 \Sigma \Delta U_\phi/t_w, 
\end{equation}
which introduces the additional parameter $\Delta U_\phi$, which is the amount by which the angular momentum of the disk gas changes over the scale height of the disk (before it is launched into the wind). 

Our governing equation therefore takes the form 
\begin{multline}
{\partial \Sigma \over \partial t} - {1\over r} {\partial \over \partial r}\left[\sqrt{4r\over GM} {\partial \over \partial r} \left({r\Sigma W^r_\phi }\right)\right] \\ =  - {2\Sigma \over t_w} + {2\over r}{\partial \over \partial r}\left(\sqrt{4r^3 \over GM} {\Sigma \over t_w} \Delta U_\phi\right) .
\end{multline}
While we shall always keep this evolutionary equation linear (in the surface density), we shall consider a range of reasonable parameterizations of the outflow physics. This includes phenomenological radius-dependent wind timescales $t_w(r)$, as well as potentially more physically natural parameterizations in terms of the local accretion rate in the disk. 

\subsection{Constructing Green's functions}
Throughout this paper we will make the simplifying assumption that both the turbulent stress $W^r_\phi$, the wind timescale $t_w$, and the angular momentum change $\Delta U_\phi$, are functions only of position in the disk, and not of the surface density $\Sigma$. This keeps the differential operator in the governing equation a linear function of the surface density $\Sigma$. This allows simple techniques to be employed to derive Greens function solutions of the evolutionary equation. 

For a linear differential operator ${\cal L}(y)$ which acts only position, the equation 
\begin{equation}
    {\partial y \over \partial t} + {\cal L}(y) = 0, 
\end{equation}
is solved by $y(x, t) = y_s(x) e^{-st}$, where $y_s(x)$ is an eigenfunction of the operator with eigenvalue $s$
\begin{equation}
    {\cal L}(y_s) = s y_s. 
\end{equation}
As our operator is a second order linear differential operator, we know that it has a continuous spectrum of eigenvalues, and that the associated eigenfunctions are orthogonal and span the full function space. This means that a general solution can be expressed as an expansion over these eigenfunctions 
\begin{equation}
    y(x, t) = \int a(s) y_s(x) e^{-st} \, {\rm d}s, 
\end{equation}
meaning that the problem becomes one of determining the modal coefficients $a(s)$ and the eigenfunctions $y_s(x)$. If we have an initial condition $y(x, t=0) = Y(x)$, then 
\begin{equation}
    Y(x) = \int a(s) y_s(x) \, {\rm d}s, 
\end{equation}
which by spatial orthogonality allows us to express 
\begin{equation}
    a(s)= \int Y(x) y_s(x)\, {\rm d}x,
\end{equation}
and so 
\begin{equation}
    y(x, t) = \int Y(x')\left[\int y_s(x') y_s(x) e^{-st} \, {\rm d}s \right]\, {\rm d}x', 
\end{equation}
from which we see that the Greens function equals
\begin{equation}
    G(x, x', t) \equiv \int y_s(x') y_s(x) e^{-st} \, {\rm d}s .
\end{equation}
Therefore our problem becomes only one of finding the eigenfunctions of the disk differential operator in the presence of various outflows. Of use throughout this paper is the general result that the solution of 
\begin{equation}
    {\partial^2 y \over \partial x^2 } - {2\alpha - 1 \over x} {\partial y \over \partial x} + \left(\beta^2 \gamma^2 x^{2\gamma-2} + {\alpha^2 - \nu^2 \gamma^2 \over x^2}\right)y = 0, 
\end{equation}
which vanishes at the origin, is \citep{Bowman58}
\begin{equation}
    y = x^\alpha J_\nu(\beta \, x^\gamma). 
\end{equation}
Then the modal superposition integral \citep{Gradshteyn80}
\begin{multline}\label{mode_sup}
\int_0^\infty  J_\nu(2X \sqrt{s}) J_\nu(2 X_0 \sqrt{s}) e^{-st} {\rm d}s \\
= {1 \over t} \exp\left({-X^2 - X_0^2 \over  t} \right) I_\nu \left({2 X X_0 \over  t}\right),
\end{multline}
provides the final form of the Greens function. 

\section{Zero-torque outflows}
As a starting case, let us consider outflows which do not torque the disk. This corresponds to matter being removed from the flow which carries with it the Keplerian angular momentum of a circular orbit, i.e., $\partial U_\phi /\partial z = 0$. This implies that $F_J = 0$. 

We will consider a general situation where both the wind time and the turbulent stress are functions of position, but not of disk density.  
In this limit we have 
\begin{equation}
    {\partial \Sigma \over \partial t} - {1\over r} {\partial \over \partial r}\left[\sqrt{4r\over GM} {\partial \over \partial r} \left({r\Sigma W^r_\phi }\right)\right]  =  - {2\Sigma \over t_{w}},
\end{equation}
we multiply through by $rW^r_\phi$ (remembering that $W^r_\phi$ is a function only of position), and define $h \equiv \sqrt{GMr}$ and $\xi \equiv r\Sigma W^r_\phi$, to leave 
\begin{equation}
    {\partial \xi \over \partial t} + {2\xi \over t_{w}} - W^r_\phi {GM \over 2h} {\partial^2 \xi \over \partial h^2} = 0. 
\end{equation}
The modal eigenfunction equation we wish to solve is 
\begin{equation}
    {\partial^2 \xi_s \over \partial h^2} = {4h \over GMW^r_\phi t_w} \xi_s -{2hs\over GMW^r_\phi} \xi_s .
\end{equation}
Comparing the above functional form to the general Bessel equation discussed above we see that simple Bessel function solutions exist to this equation when (i) the stress $W^r_\phi$ depends on the radius to some general power, and (ii) the product 
\begin{equation}
    {4h \over GMW^r_\phi t_w} = {A \over h^2}, 
\end{equation}
where $A$ is a constant. This is in effect a naturalness condition, as the viscous time $(t_{\rm visc} \equiv {h^3 / {GMW^r_\phi}})$ and wind timescales must be proportional 
\begin{equation}
    t_{\rm visc}\propto t_w ,
\end{equation}
for simple solutions to be found. For winds launched via magnetic mechanisms this may well in fact be relatively natural, as the same magnetic fields which redistribute angular momentum radially (setting the ``viscous'' timescale) will also be launching material vertically (setting the ``wind'' timescale). 

For an explicit example let us take $W^r_\phi = w = {\rm constant}$ (itself rather natural, as in this limit the viscous time is proportional to the orbital time), and therefore a wind time proportional to the orbital time 
\begin{equation}
    t_w = t_{w, 0} (h/h_0)^3, 
\end{equation}
where $t_{w, 0}$ and $h_0$ are the wind timescale at a reference radius, $h_0 = \sqrt{GMr_0}$. One is not formally free to chose just any amplitude of $t_{w, 0}$, as the governing thin disk equation is derived on the premise that all velocities in the problem are sub dominant to the orbital velocity. Or in effect, we require $t_{w, 0} \gg t_{\rm orb, 0}$, where $t_{\rm orb, 0}$ is the orbital time at the reference radius. (This is an identical mathematical condition to the requirement of long viscous times for a self consistent thin disk theory in the absence of outflows.)

Then, further defining $\varphi \equiv h/ h_0$, $\tilde s \equiv 2h_0^3 s/GMw$ and $\tilde t \equiv GMw t_{w, 0}/2h_0^3$, we have the modal equation 
\begin{equation}
    {\partial^2 \xi_s \over \partial \varphi^2} + \left[\tilde s \varphi - {2\over \tilde t} {1\over \varphi^2}\right] \xi_s = 0. 
\end{equation}
The eigenfunction solution for this equation is 
\begin{equation}
    \xi_s(\varphi) = \varphi^{1/2}  J_{\nu} \left({2\over 3 }\sqrt{\tilde s} \, \varphi^{3/2}\right), 
\end{equation}
where, crucially 
\begin{equation}
    \nu \equiv {1\over 3} \sqrt{1 + {8 \over \tilde t}}. 
\end{equation}
This is a remarkably simple solution of this coupled inflow-outflow problem. At first it appears that not much has changed in the physical character of this solution (when compared to the classical windless solutions). This is however an illusion of this compact notation, the change in Bessel index $\nu$ carries with it significant implications for the disk structure. 

To highlight some of these implications let us construct the explicit Green's function (for the combination $\xi$), which is given for these modal solutions  by 
\begin{equation}
    G(r, r_0, t) \propto {r^{1/4} \over t} \exp\left({-1 - (r/r_0)^{3/2} \over 4t/t_{v, 0}}\right) I_\nu\left({(r/r_0)^{3/4} \over 2t/t_{v, 0}}\right) .
\end{equation}
Here we have defined the ``viscous'' timescale $t_{v, 0}$ at the initial radius of the flow $r_0$ by $t_{v,0} \equiv {2\sqrt{GMr_0^3}/9w}$. 

Most of the interesting physics of these Green's functions is contained within the small-radius large-time expansion of the Bessel function, explicitly 
\begin{equation}
    \lim_{z\to 0}I_\nu(z) \sim z^\nu, 
\end{equation}
such that 
\begin{equation}
    G \sim r^\chi t^{-n} ,
\end{equation}
where 
\begin{align}
    \chi \equiv {1\over 4} + {1\over 4}\sqrt{1 + 72{t_{v, 0}\over t_{w, 0}}},\\
    n \equiv 1 + {1\over 3} \sqrt{1 + 72{t_{v, 0}\over t_{w, 0}}}. 
\end{align}
We see that, depending on the ratio of the natural wind and viscous timescales, these Green's function solutions can differ markedly from their windless ($t_{w, 0}\to \infty$) cousins. 

To see this explicitly, note that the temperature profile of these disks $T^4 \propto \xi r^{-7/2}$, such that the temperature index is a sensitive function of timescales
\begin{equation}
    \log (T) \propto -\left({13\over 16} - {1\over 16}\sqrt{1 + 72 {t_{v, 0}\over t_{w, 0}}}\,\right) \log r. 
\end{equation}
In other words, winds (with short timescale compared to the viscous time) cause a significant flattening of the disk temperature profile. The bolometric luminosity of the disk is found from integrating $T^4$ over the disk area, and is dominated by the $r\to 0$ limit. The bolometric luminosity falls at large times as 
\begin{equation}
    \log (L_{\rm bol}) \propto - \left(1 + {1\over 3} \sqrt{1 + 72{t_{v, 0}\over t_{w, 0}}}\,\right) \log(t) .
\end{equation}
This (for short wind timescales) represents a significant steepening of the bolometric decay index. 

It is often more natural, for certain problems, to work with the Green's function for the accretion rate itself $\dot M_{\rm acc}(t)$, or for the outflowing mass rate $\dot M_{\rm out}(t)$. In the limit of zero torque from the wind the mass accretion rate is given by 
\begin{equation}
    \dot M_{\rm acc}(r, t) = - 2\pi \sqrt{4r \over GM} {\partial \xi \over \partial r}, 
\end{equation}
and so our mass accretion rate Green's function (for constant $W^r_\phi$, a result generalized and normalized in the Appendix) is 
\begin{multline}
    G_{\dot M_{\rm acc}}(r, r_0, t) \propto {1\over t} \left({r\over r_0}\right)^{-1/4} \exp\left({-1 - (r/r_0)^{3/2} \over 4t_/t_{v, 0}}\right) \\ 
    \Bigg[{3\over 8} \left({t_{v, 0}\over t}\right)\left({r\over r_{0}}\right)^{3/4} I_{\nu-1}\left({(r/r_0)^{3/4} \over 2t/t_{v, 0}}\right) \\ + \left({1\over 4}-{3\nu \over 4} - {3\over 8} \left({t_{v, 0}\over t}\right)\left({r\over r_{0}}\right)^{3/2} \right)I_{\nu}\left({(r/r_0)^{3/4} \over 2t/t_{v, 0}}\right)  \Bigg] .
\end{multline}
An interesting difference between this windy Green's function and the windless Green's functions of \cite{LBP74} is that the mass accretion rate at the origin ($r\to 0$) is zero, as opposed to converging to a finite value. Indeed, the small radius scaling is found from repeat use of $I_k(z) \sim z^k$ for small $z$, leading to 
\begin{multline}
    \log \dot M_{\rm acc}(r\ll r_0, t) \propto \left(-{1\over 4} + {1\over 4}\sqrt{1 + 72 {t_{v, 0}\over t_{w, 0}}}\right) \log r \\ - \left(1 + {1\over 3} \sqrt{1 + 72{t_{v, 0}\over t_{w, 0}}}\,\right) \log(t) 
\end{multline}
We note that the decay of the bolometric luminosity still follows the decay of the accretion rate at the origin, even in the presence of outflows. 

The Green's function for the mass outflow rate from all radii interior to radius $r$ is then given by
\begin{equation}
    G_{\dot M_{\rm out}}(r, r_0, t) = \int^r 4\pi r' G_\Sigma(r', r_0, t) {1\over t_w(r')}\, {\rm d}r', 
\end{equation}
with the integrand giving the local mass outflow rate. This integral is finite at the lower limit (despite the fact that the outflow rate is  poorly defined as $r\to0$), and is always of the form 
\begin{equation}
    G_{\dot M_{\rm out}}(t) \propto \int^r {x^{-5/4} \over t} \exp\left({-1-x^{3/2} \over 4 t/t_{v, 0}}\right) I_\nu \left({x^{3/4}\over 2 t/t_{v, 0}}\right)\, {\rm d}x. 
\end{equation}
To see that this is finite as $r\to 0$, use the Bessel function expansion to find 
\begin{multline}
    \log \dot M_{\rm out}(r\ll r_0, t) \propto \left(-{1\over 4} + {1\over 4}\sqrt{1 + 72 {t_{v, 0}\over t_{w, 0}}}\right) \log r \\ - \left(1 + {1\over 3} \sqrt{1 + 72{t_{v, 0}\over t_{w, 0}}}\,\right) \log(t) ,
\end{multline}
of precisely the same scaling form as the local accretion rate. Indeed, this is a generic property of all of the solutions derived in this work, a result which can be readily understood. Mass conservation in these solutions is given by 
\begin{equation}
    {\partial \Sigma \over \partial t} + {1\over r}{\partial \over \partial r}(r\Sigma U^r) = - F_M, 
\end{equation}
Meaning that at late times (when small $s$ modes dominate and $\partial_t\approx 0$) we can integrate these expression from $0$ to $r$ to find
\begin{equation}
    \left[2\pi r \Sigma U^r\right](r, t) \approx -\int_0^r 2\pi r' F_M(r')\, {\rm d}r',
\end{equation}
or explicitly 
\begin{equation}
    |\dot M_{\rm acc}(r, t)| \approx |\dot M_{\rm out}(<r, t)|, 
\end{equation}
where we use the notation $<r$ to indicate the mass outflow at all radii smaller than $r$ (note that this argument is independent of any torques generated by disk winds). 

As we shall show, all of the Green's function solutions to the Newtonian thin disk evolution equation with outflows have qualitatively identical behavior (although the quantitative details of course depend on the choice of wind parameterisation). As such, we delay a detailed discussion of the properties of these solutions (and plots of these solutions) until section 7.

All of the above results can be trivially extended to more general stress profiles $W^r_\phi \propto r^\mu$, provided that we enforce the naturalness condition $t_w \propto r^{3/2-\mu}$. This is performed in the Appendix. 

\section{Torquing outflows }
The above analysis assumed that the outflowing material did not impart a torque on the flow, or in other words the outflowing material carried with it it's Keplerian angular momentum. 

Let us now relax that assumption, and take 
\begin{equation}
    \Delta U_\phi = \Gamma  U_\phi, 
\end{equation}
with the as of yet unspecified (but assumed constant) parameter $\Gamma$. Note that a purely radial wind (i.e., the flow returns all of its angular momentum to the disk on its way through the atmosphere) is given by $\Gamma = -1$, and we have generically that $\Gamma \geq -1$. In principle a very small amount of material can carry a large angular momentum budget, and so $\Gamma$ is not formally bounded from above. 

Substituting this into our governing equation, we find 
\begin{multline}
{\partial \Sigma \over \partial t} - {1\over r} {\partial \over \partial r}\left[\sqrt{4r\over GM} {\partial \over \partial r} \left({r\Sigma W^r_\phi }\right)\right] \\ =  - {2\Sigma \over t_w} + {4\Gamma\over r}{\partial \over \partial r} \left( r^2 {\Sigma \over t_w}\right) .
\end{multline}
Again, defining $\xi \equiv r \Sigma W^r_\phi$, $h =\sqrt{GMr}$, we find 
\begin{multline}
    {\partial \xi \over \partial t} - W^r_\phi {GM\over 2h} {\partial ^2 \xi \over \partial h^2} \\ = - {2\xi \over t_w} + {4\Gamma}W^r_\phi {GM\over 2h} {\partial \over \partial h}\left({h^2 \over GMW^r_\phi t_w} \xi\right) . 
\end{multline}
Our eigenvalue problem is therefore expressed as 
\begin{multline}
    {\partial ^2 \xi_s \over \partial h^2} + {4\Gamma h^2 \over GMW^r_\phi t_w}  {\partial \xi_s \over \partial h} = {4h \over GMW^r_\phi t_w}\xi_s - {2h s\over GMW^r_\phi} \xi_s \\ - 4\Gamma \left[{\partial \over \partial h}\left({h^2 \over GMW^r_\phi t_w}\right) \right]\xi_s .
\end{multline}
We again find the exact same naturalness condition for the presence of simple analytical solutions, i.e., 
\begin{equation}
    {h^2 \over  W^r_\phi t_w} \propto {1\over h},
\end{equation}
or equivalently $t_w \propto t_{\rm visc}$. As an explicit example we once again take $W^r_\phi = w = {\rm constant}$, $t_w = t_{w, 0} (h/h_0)^3$ and define $\varphi \equiv h/ h_0$, $\tilde s \equiv 2h_0^3 s/GMw$ and $\tilde t \equiv GMw t_{w, 0}/2h_0^3$. This leaves the modal equation 
\begin{equation}
    {\partial ^2 \xi_s \over \partial \varphi^2} + {2\Gamma \over \tilde t} {1\over \varphi} {\partial \xi_s \over \partial \varphi}+ \left[\tilde s \varphi - {2(1+\Gamma)\over \tilde t} {1\over \varphi^2}\right] \xi_s = 0. 
\end{equation}
This again has simple Bessel function solutions of the form 
\begin{equation}
    \xi_s(\varphi) = \varphi^{1/2-\Gamma/\tilde t}  J_{\nu} \left({2\over 3 }\sqrt{\tilde s} \varphi^{3/2}\right), 
\end{equation}
where, now,  
\begin{equation}
    \nu \equiv {1\over 3} \sqrt{1 + {4\Gamma^2 + 4 \Gamma \tilde t+ 8\tilde t \over \tilde t^2}}. 
\end{equation}
Going from these eigenfunction solutions to the full Green's function is trivial, and follows an identical procedure to above.

We see that, once again, the presence of an outflow (even one which torques the flow) results in apparently minor mathematical changes to the formal Green's function solutions of the disk equations. The physical content of a torquing wind is effectively identical to that of the torque-less limit, only with the slight modification of $8/\tilde t \to (4\Gamma^2 + 4 \Gamma \tilde t+ 8\tilde t)/ \tilde t^2$ and $1/2 \to 1/2-\Gamma/\tilde t$ in various indices. A more detailed discussion of the mathematical properties of these solutions is presented in the Appendix. 

\section{Inflow-outflow coupling}
For accretion flows accreting at high ($\gtrsim$ Eddington) rates, it is perhaps more natural for the outflowing mass rate to be coupled to the local accretion rate, rather than just to the local density multiplied by a radius-dependent timescale. Indeed, we derived in an earlier section that this is a generic asymptotic property of all such evolutionary solutions in the large time limit.  In this section we examine the properties of such flows. 

We will, for simplicity, neglect any torque on the flow (i.e., $F_J = 0$). While this is a choice of convenience, we demonstrated in the previous section that the impacts of such torques are quantitative, rather than resulting in qualitative changes from the torque-less limit. 

The local mass accretion rate in the torque-less limit is given by 
\begin{equation}
    \dot M_{\rm acc}(r, t) = 2\pi r U^r \Sigma = - 2\pi \sqrt{4r\over GM} {\partial \over \partial r} (r \Sigma W^r_\phi). 
\end{equation}
Setting the outgoing mass flux equal to 
\begin{equation}
    F_M = -\varepsilon_w {\dot M_{\rm acc}(r, t)\over \pi r^2 } ,  
\end{equation}
where $\varepsilon_w$ is a coefficient of proportionality\footnote{A better model, particularly at early times, would have an absolute value $|\dot M_{\rm acc}(r, t)|$, rather than $-\dot M_{\rm acc}(r, t)$, however this non-linear equation escapes a simple analysis.} is therefore a physically rather natural parameterisation. If $F_M$ where a constant with radius then $\varepsilon_w$ would just represent the ``wind efficiency'' -- namely the ratio of the mass launched in outflows interior to $r$ to the local accretion rate at $r$. One requires $\varepsilon_w\geq0$, but we note that $\varepsilon_w>1$ is formally allowed (and may even be likely in the super-Eddington limit). 

Recalling that $\dot M = -2\pi r U^r \Sigma$, and that $F_M = 2 \rho U^z$, we note that an equivalent interpretation of $\varepsilon_w$ can be obtained from combining the above two expressions to find  
\begin{equation}
    U^z = - \varepsilon_w \left({H\over r}\right) U^r. 
\end{equation}
For $\varepsilon_w \lesssim 1$ and $H/r \lesssim 1$ the asymptotic formalism under which the governing equations are derived remains robustly valid. 

\begin{figure*}
    \centering
    \includegraphics[width=0.49\linewidth]{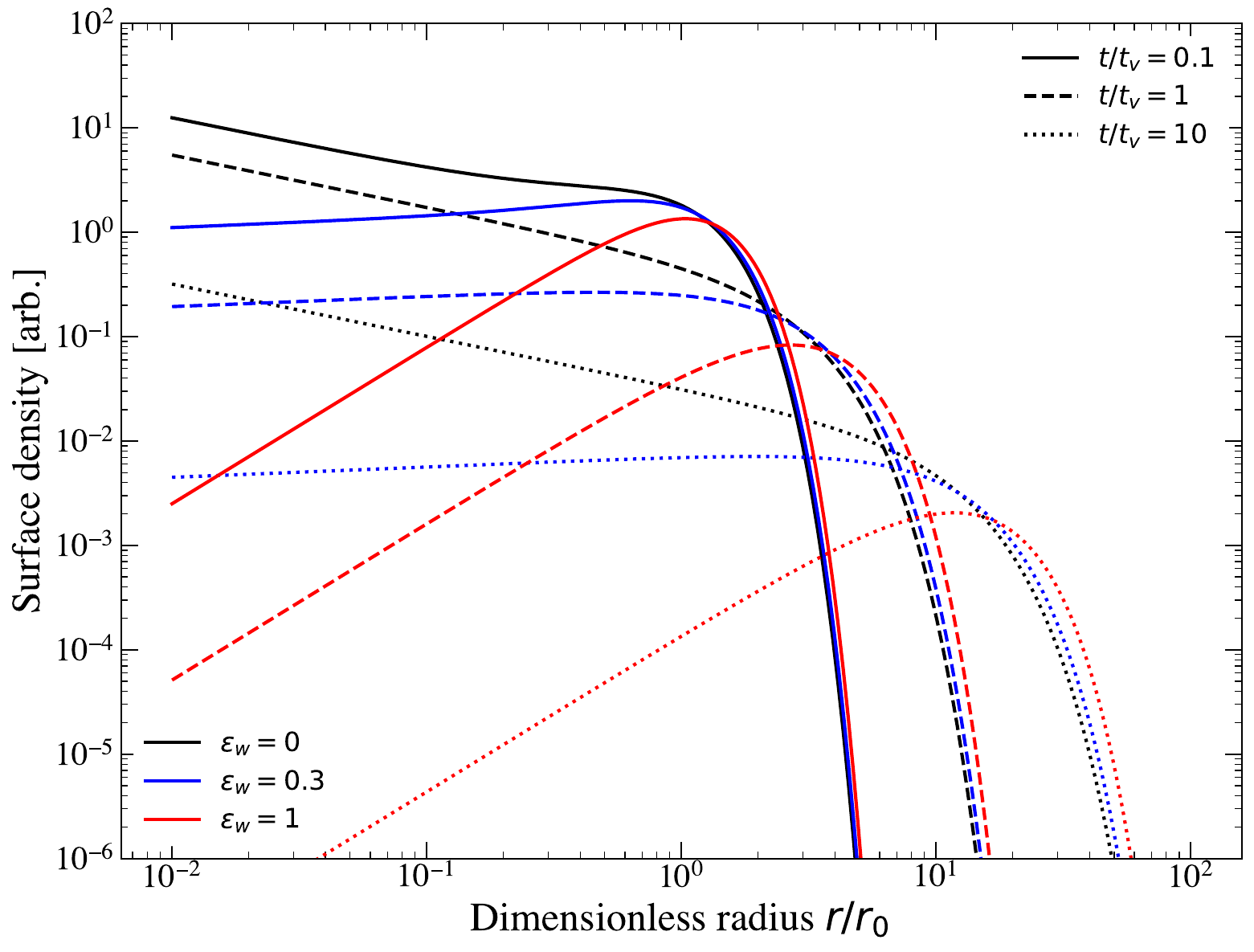}
    \includegraphics[width=0.49\linewidth]{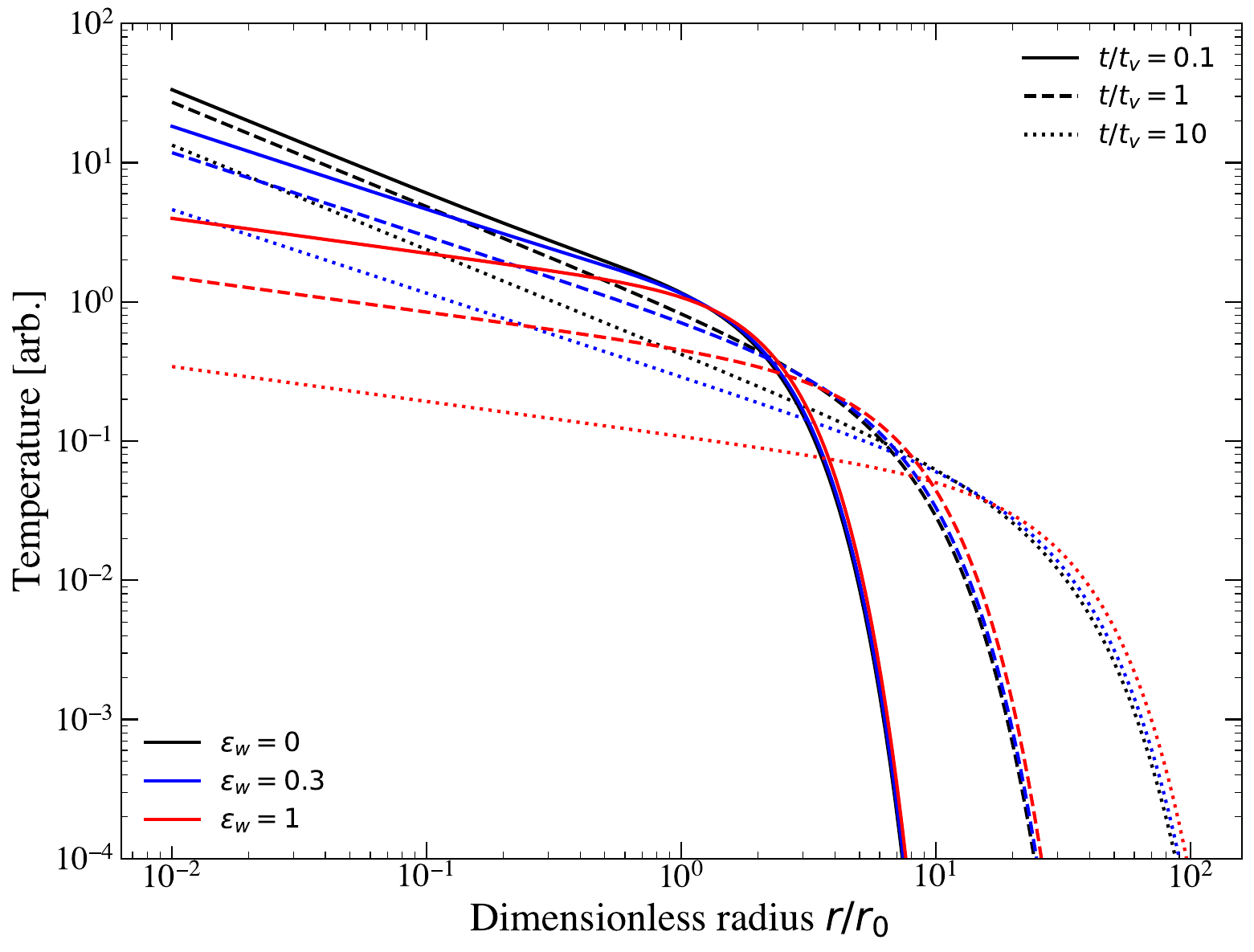}
    \caption{{\bf Left:} the evolving surface density (arbitrarily but consistently normalised) of Green's function solutions to the constant $\varepsilon_w$ coupled inflow-outflow disk evolution equation, at three different dimensionless times (denoted by line style) and three different outflow parameters $\varepsilon_w$ (denoted by color). Stronger outflows strongly suppress the surface density of the inner disk regions. The evolution of the outer disk regions is almost unaffected by the outflows. {\bf Right:} the evolving disk temperature of the same disk solutions. Outflows flatten the disk temperature profile markedly, and strongly suppress the peak value of the inner disk temperature. }
    \label{fig:dens+temp}
\end{figure*}

Substituting this parameterisation into the governing equation leads to 
\begin{multline}
    {\partial \Sigma \over \partial t} - {1\over r} {\partial \over \partial r}\left[\sqrt{4r\over GM}{\partial \over \partial r}(r\Sigma W^r_\phi)\right] \\ = -{2\varepsilon_w\over r^2} \sqrt{4r\over GM} {\partial \over \partial r} (r \Sigma W^r_\phi).
\end{multline}
Multiplying by $r W^r_\phi$, and again moving to $h = \sqrt{GMr}$ leads to our formal eigenfunction problem
\begin{equation}
    {\partial^2 \xi_s \over \partial h^2} - {4\varepsilon_w \over h} {\partial \xi_s \over \partial h} + {2h s\over GMW^r_\phi} \xi_s = 0. 
\end{equation}
For any turbulent stress with a power-law dependence on radius, this is of the form of the Bessel equation. 
By again moving to $\varphi \equiv h/h_0$, with $\tilde s = {2h_0^3 s/GMw}$ and $W^r_\phi = w = {\rm constant}$ (a final assumption trivially generalized), we have 
\begin{equation}
    \xi_s(\varphi) = \varphi^{1/2 + 2\varepsilon_w} J_\nu\left({2\over 3 }\sqrt{\tilde s} \varphi^{3/2}\right), 
\end{equation}
where now 
\begin{equation}
    \nu \equiv {1\over 3} + {4\over 3}\varepsilon_w. 
\end{equation}
We again have a Green's function of the same general form, namely 
\begin{equation}
    G(r, r_0, t) \propto     {r^{{1\over 4}+\varepsilon_w} \over t} \exp\left({-1 - (r/r_0)^{3/2} \over 4t/t_{v, 0}}\right) I_\nu\left({(r/r_0)^{3/4} \over 2t/t_{v, 0}}\right) .
\end{equation}
Again, we see the clear impacts of an outflow on the disk structure, the temperature profile of the disk is flattened (see Figure \ref{fig:dens+temp}), where now 
\begin{equation}
    \log(T) \propto -\left({3\over 4} - {1 \over 2}\varepsilon_w \right) \log r ,
\end{equation}
and bolometric decay index 
\begin{equation}
    \log(L_{\rm bol}) \propto - \left({4\over 3} + {4 \over 3}\varepsilon_w \right) \log(t). 
\end{equation}
The Green's function of the accretion rate can also be written down explicitly, and is proportional to 
\begin{multline}
    G_{\dot M_{\rm acc}}(r, r_0, t) \propto {1\over t} \left({r\over r_0}\right)^{-{1\over4} + \varepsilon_w} \exp\left({-1 - (r/r_0)^{3/2} \over 4t_/t_{v, 0}}\right) \\ 
    \Bigg[{3\over 8} \left({t_{v, 0}\over t}\right)\left({r\over r_{0}}\right)^{3/4} I_{\nu-1}\left({(r/r_0)^{3/4} \over 2t/t_{v, 0}}\right) \\ + \left({1\over 4}+\varepsilon_w-{3\nu \over 4} - {3\over 8} \left({t_{v, 0}\over t}\right)\left({r\over r_{0}}\right)^{3/2} \right)I_{\nu}\left({(r/r_0)^{3/4} \over 2t/t_{v, 0}}\right)  \Bigg] .
\end{multline}
The small radius scaling of this Green's function is 
\begin{multline}
    \log \dot M_{\rm acc}(r\ll r_0, t) \propto 2\varepsilon_w \log r \\ - \left({4\over 3} + {4\over 3}\varepsilon_w\,\right) \log(t) .
\end{multline}
The mass outflow Green's function is 
\begin{equation}
    G_{\dot M_{\rm out}} = \int^r {2\varepsilon_w \over r'}  G_{\dot M_{\rm acc}}(r', r_0, t) \, {\rm d}r',
\end{equation}
which again cannot be performed analytically, but has a small radius scaling identical to the accretion rate 
\begin{multline}
    \log \dot M_{\rm out}(r, t) \propto 2\varepsilon_w \log r \\ - \left({4\over 3} + {4\over 3}\varepsilon_w\,\right) \log(t) .
\end{multline}
Again, this is merely a result of mass conservation in the large time asymptotic limit. 
\section{Variability and the Fourier-Green function}
The disk evolution equation, while normally used to study large scale evolution of the flow, also describes the evolution of individual perturbations in the disk density. It therefore provides a useful framework to examine the turbulent variability in various disk properties (a framework generally referred to as the ``theory of propagating fluctuations'', \citealt{Lyubarskii97, Mushtukov2018, Mummery23b}). 

Practically it is most useful to examine the properties of variability in the Fourier domain, and therefore the Fourier transform of the various Green's functions derived in this work are of interest. In this section we drop dimensional parameters such as $GM, c, w,$ etc. in our analysis, moving to dimensionless radii $x$ and $x_0$.  The quantities we compute (for example the Fourier transform of the accretion rate $\widetilde G_{\dot M}(x, x_0, f)$) correspond physically to the variability at frequency $f$ in a disk quantity at radius $x$ sourced by a perturbation to the disk at radius $x_0$. 

The Fourier transform of $G(x, x_0, t)$, denoted $\widetilde G(x, x_0, f)$ is defined by  the complex integral
\begin{equation}
\widetilde G(x, x_0, f) \equiv \int_0^\infty G(x, x_0, t) \exp(-2\pi i f t) \, {\rm d} t ,
\end{equation}
where we have used the fact that $G(x, x_0, t<0) = 0$. When written in terms of the eigenfunction superposition, this integral becomes (we keep our notation general here, defining $q(x)$ and $g(x)$ which vary for the various different wind parameterizations functions)
\begin{multline}
\widetilde G(x, x_0, f) = q(x)\int_0^\infty \Bigg[ \int_0^\infty  J_\nu(\sqrt{s} g(x)) J_\nu(\sqrt{s} g(x_0)) \\ \exp({-st}) \, {\rm d}s  \Bigg]  \exp(-2\pi i f t) \, {\rm d} t .
\end{multline}
As both integrals converge, we can  swap the order of integration 
\begin{multline}
\widetilde G(x, x_0, f) = q(x)\int_0^\infty \Bigg[ \int_0^\infty  \exp(-st - 2\pi i ft) \, {\rm d}t \Bigg] \\  J_\nu(\sqrt{s} g(x)) J_\nu(\sqrt{s} g(x_0))  \, {\rm d}s  
\end{multline}
which is more easily solved. Performing the $t$ integral leaves
\begin{equation}
\widetilde G(x, x_0, f) = q(x) \int_0^\infty { J_\nu(\sqrt{s} g(x)) J_\nu(\sqrt{s} g(x_0)) \over s + 2\pi i f} \, {\rm d}s  .
\end{equation}
By making the substitution $u = \sqrt{s}$, this integral becomes 
\begin{equation}
\widetilde G(x, x_0, f) = 2 q(x) \int_0^\infty { u J_\nu(u g(x)) J_\nu( u g(x_0)) \over u^2 + \beta^2} \, {\rm d}u  ,
\end{equation}
where 
\begin{equation}
\beta \equiv (1 + i) \sqrt{\pi f}.
\end{equation}
When written in this form the solution of the integral is a standard result, which can be found in the text of \cite{Gradshteyn80}
\begin{equation}
\widetilde G(x, x_0, f) = 2 q(x) 
\begin{cases}
& I_\nu(\beta g(x)) K_\nu(\beta g(x_0)) , \quad x< x_0, \\
\\
& I_\nu(\beta g(x_0)) K_\nu(\beta g(x)) , \quad x> x_0 .
\end{cases}
\end{equation}  
In this expression $K_\nu$ is the modified Bessel function of the second kind.   The Green's function solutions for the mass accretion rate, denoted $G_{\dot M}$, are also of interest, as it is often assumed that variability in the mass accretion rate is directly communicated into variability in the locally emitted flux. 
The mass accretion rate Green's function has the following form
\begin{equation}
G_{\dot M} (x, x_0, t) = x^{1/2} {\partial \over \partial x} G(x, x_0, t) .
\end{equation}
In the Fourier domain (and specializing to the torque-free wind limit)
\begin{align}
\widetilde G_{\dot M} (x, x_0, f) &\equiv \int_0^\infty G_{\dot M}(x, x_0, t) \exp(-2\pi i f t) \, {\rm d} t,  \nonumber \\ 
&=   \int_0^\infty x^{1/2} {\partial \over \partial x} \left[G(x, x_0, t)\right] \exp(-2\pi i f t) \, {\rm d} t, \nonumber \\ 
&=  x^{1/2} {\partial \over \partial x}\widetilde G(x, x_0, f),
\end{align}
where in going to the final line we have used the fact that the $x$ derivative and $t$ integral commute. The physical interpretation of $\widetilde G_{\dot M}$ is the variability (at frequency $f$) in the accretion rate at disk radius $x$ from a perturbation sourced at radius $x_0$. 

We therefore have the general solution 
\begin{equation}\label{general_def}
\widetilde  G_{\dot M} \propto  x^{1/2} 
\begin{cases}
&  K_\nu(\beta g(x_0)) \,  {\partial_x} \left[ q(x) I_\nu(\beta g(x))\right] , \quad x< x_0, \\
\\
& I_\nu(\beta g(x_0)) \,  {\partial_x} \left[ q(x)  K_\nu(\beta g(x)) \right] , \quad x> x_0 ,
\end{cases}
\end{equation}  
where we use the notation $\partial_x \equiv \partial/\partial x$. 

To be concrete in the impact of outflows on the variability properties of the mass accretion rate, move to the specific solution for $W^r_\phi = w ={\rm constant}$, and a constant inflow-outflow efficiency $\varepsilon_w$. In this limit $q(x) = x^{1/4 + \varepsilon_w}$, $g(x) = Ax^{3/4}$ (for constant $A\propto x_0^{1/4}$), and $\nu = 1/3 + 4\varepsilon/3$. In this limit the Fourier-Greens functions have a compact final form
\begin{equation}\label{eff_def}
\widetilde  G_{\dot M} \propto  
\begin{cases}
& +A\beta x^{{1\over 2}+\varepsilon_w}  K_{{1\over 3}+{4\over 3}\varepsilon_w}(A\beta x_0^{3/4}) \,  I_{-{2\over 3} + {4\over 3}\varepsilon_w}(A\beta x^{3/4}) , \\ &\quad\quad\quad\quad\quad\quad\quad\quad\quad\quad\quad\quad (x< x_0), \\
\\
& -A\beta x^{{1\over 2}+\varepsilon_w}  I_{{1\over 3}+{4\over 3}\varepsilon_w}(A\beta x_0^{3/4}) \,  K_{-{2\over 3} + {4\over 3}\varepsilon_w}(A\beta x^{3/4}), \\ &\quad\quad\quad\quad\quad\quad\quad\quad\quad\quad\quad\quad (x>x_0) .
\end{cases}
\end{equation}
While these expressions look unwieldy, they contain interesting physical properties of these solutions. Perhaps most relevantly, small radii and small frequency variability is suppressed in the presence of outflows. 

To see this explicitly, take the $x\to 0$ limit of the $x<x_0$ Fourier-Green's function, leading to 
\begin{equation}
    \widetilde G(x\to 0, x_0, f) \propto x^{2\varepsilon_w} \beta^{{1\over 3} +{4\over 3}\varepsilon_w} K_{{1\over 3}+{4\over 3}\varepsilon_w}(A\beta x_0^{3/4}).
\end{equation}
We see that in the wind-less limit ($\varepsilon_w=0$) the variability is a constant as $x\to 0$, depending only on how far into the disk the perturbation was sourced and the frequency. However, in the presence of outflows, the variability at small radii is suppressed, with a suppression index given by the outflow efficiency. This can be readily understood physically, as variability in the accretion rate at small radii is necessarily reduced by the mass removed from the accretion flow by winds launched at large radii. 

Variability in the accretion rate is also suppressed at low frequencies (on long timescales). Taking the low frequency limit of these Fourier-Green's functions (i.e., the limit $\beta \to 0$, and using $K_\nu(z) \sim z^{-\nu}$ for small $z$), we find (for inwards propagation, $x<x_0$)
\begin{equation}
    \widetilde G_{\dot M}(x, x_0, f\to 0) \propto x^{2\varepsilon_w} \, x_0^{-\varepsilon_w} .
\end{equation}
We see that long-timescale variability is suppressed both for small $x$ (at fixed $x_0$) and at large perturbation source radii $x_0$ (at fixed $x$). Again, this is simple to understand physically, as the zero frequency limit of the Fourier-Green's function is just 
\begin{equation}
    \widetilde G_{\dot M}(x, x_0, f\to 0)  = \int_0^\infty G_{\dot M} (x, x_0, t)\, {\rm d}t,
\end{equation}
which, in the absence of outflows, is just the initial disk mass $M_d$ (for $x<x_0$, as all matter in the absence of winds is eventually accreted). However, when there are outflows, by definition not all of the matter is eventually accreted (as some is expelled from the system). This means that the variability amplitude is suppressed by a factor $M_{\rm out}(x, x_0)/M_d$ in this zero frequency limit (where $M_{\rm out}$ is the amount of mass expelled between $x_0$ and $x$ during the disk evolution). This naturally explains the twin scalings seen above. 

We note that the high frequency behavior of these Fourier-Green's functions is unaffected by outflows, and takes the form 
\begin{equation}
    \widetilde G_{\dot M}(x, x_0, f\to\infty) \sim \exp\left(-\sqrt{\pi f}(1+i)A\left|x^{3/4}-x_0^{3/4}\right|\right).
\end{equation}
This is a generic property of any system which on the shortest timescales (highest frequencies) has behavior dominated by diffusive operators. In other words, high frequency variability in accretion flows is damped by the disk turbulence itself.

\section{Discussion and observational implications}

In this work we have derived Green's function solutions of the Newtonian thin disk evolution equations in the presence of outflows. We have sought to be as general in our outflow profiles as possible, while maintaining analytical tractability. Analytical tractability demands (for exact solutions) linear evolutionary equations, and so we have limited the various diffusive and outflow operators in the problem to be linear functions of the disk density (or disk density derivative) only. 

We have found two analytically tractable classes of solutions. The first is where one picks a power-law parameterisation of the turbulent stress $W^r_\phi$, and then chooses a timescale for mass outflow $t_w$ directly proportional to the associated viscous time of the disk. This is in the spirit of the classical \cite{LBP74} solutions, where one is free to pick a certain ``viscosity'' parameterisation, for which general solutions can be found. The requirement $t_w \propto t_v$ is by no means required mathematically or physically for solutions to exist, it is only required for simple analytical solutions to be written down. We note that  while this timescale requirement may seem contrived, it is certainly plausible if magnetic fields are driving both the turbulent (radial) redistribution of angular momentum, as well as providing the driving mechanism for the outflows. 

The second class of solutions we have derived are in many ways more natural. Any Newtonian accretion flow with outflows will, in the long-time limit, settle into a quasi-stationary state which satisfies 
\begin{equation}
    |\dot M_{\rm acc}(r, t)| \approx |\dot M_{\rm out}(<r, t)|, 
\end{equation}
where $\dot M_{\rm acc}$ is the local accretion rate at radius $r$ and time $t$, and $\dot M_{\rm out}(<r, t)$ is the mass loss from outflows at time $t$ sourced from all radii smaller than $r$. This is a generic property of mass conservation. 

Perhaps the most natural parameterisation of the outflow is therefore through a wind efficiency $\varepsilon_w$ 
\begin{equation}
    \varepsilon_w \equiv -{\pi r^2 F_M(r, t) \over \dot M_{\rm acc}(r, t) }, 
\end{equation}
where $F_M \equiv \dot \Sigma$ is the rate at which surface density is lost from the flow (locally). Equivalently $\varepsilon_w$ can be thought of as the coupling between the inflow and outflow velocities 
\begin{equation}
    U^z = - \varepsilon_w \left({H \over r}\right) U^r. 
\end{equation}

Simple analytical solutions to the thin disk equations exist in the limit when $\varepsilon_w$ is a global constant. As we believe this is a natural framework to examine, and as the {\it qualitative} properties of these solutions are the same as for the parameterized-wind-time solutions, we now discuss the observational implications of these solutions within the framework of a constant $\varepsilon_w$. 

\subsection{The bolometric luminosity}

\begin{figure*}
    \centering
    \includegraphics[width=0.49\linewidth]{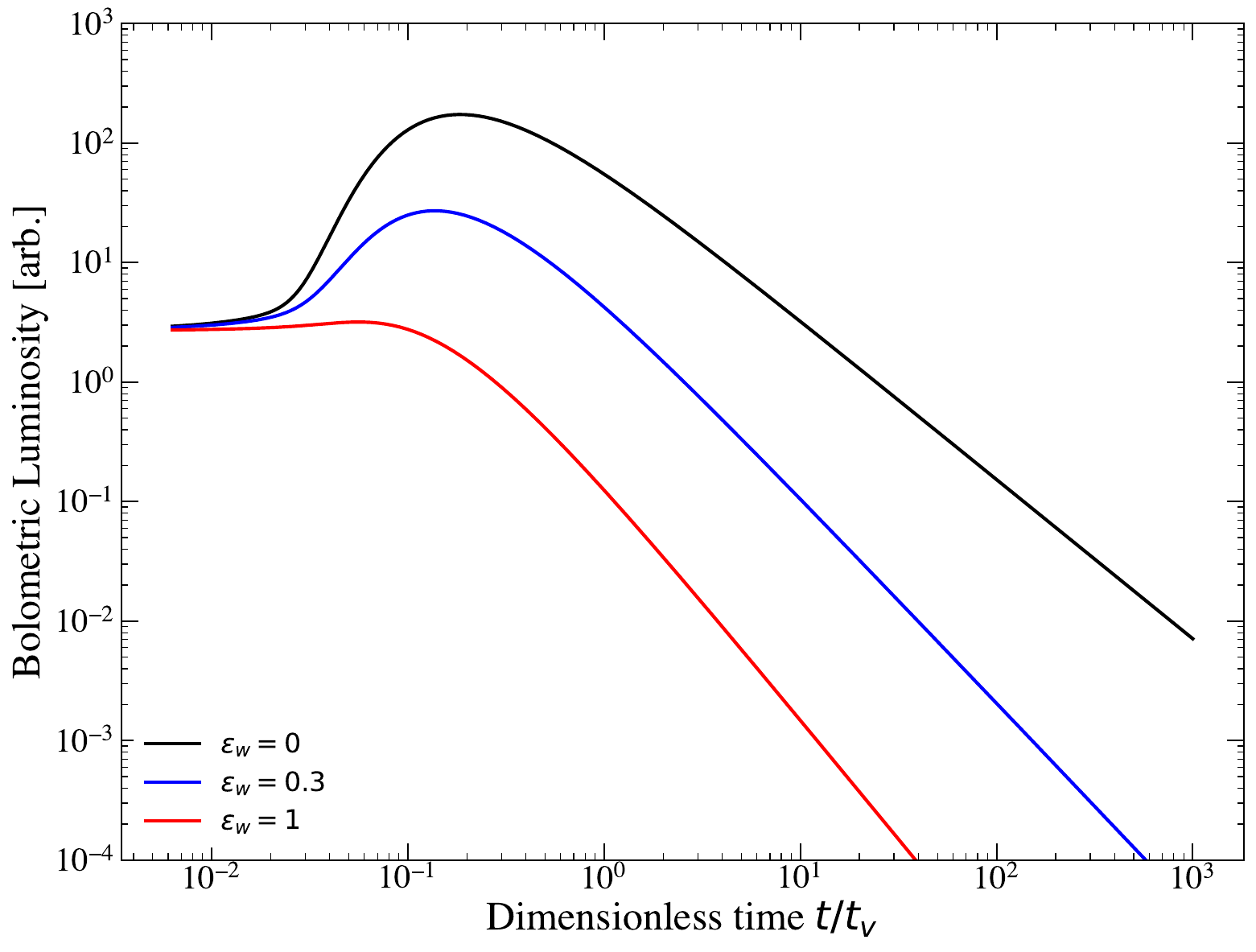}
    \includegraphics[width=0.49\linewidth]{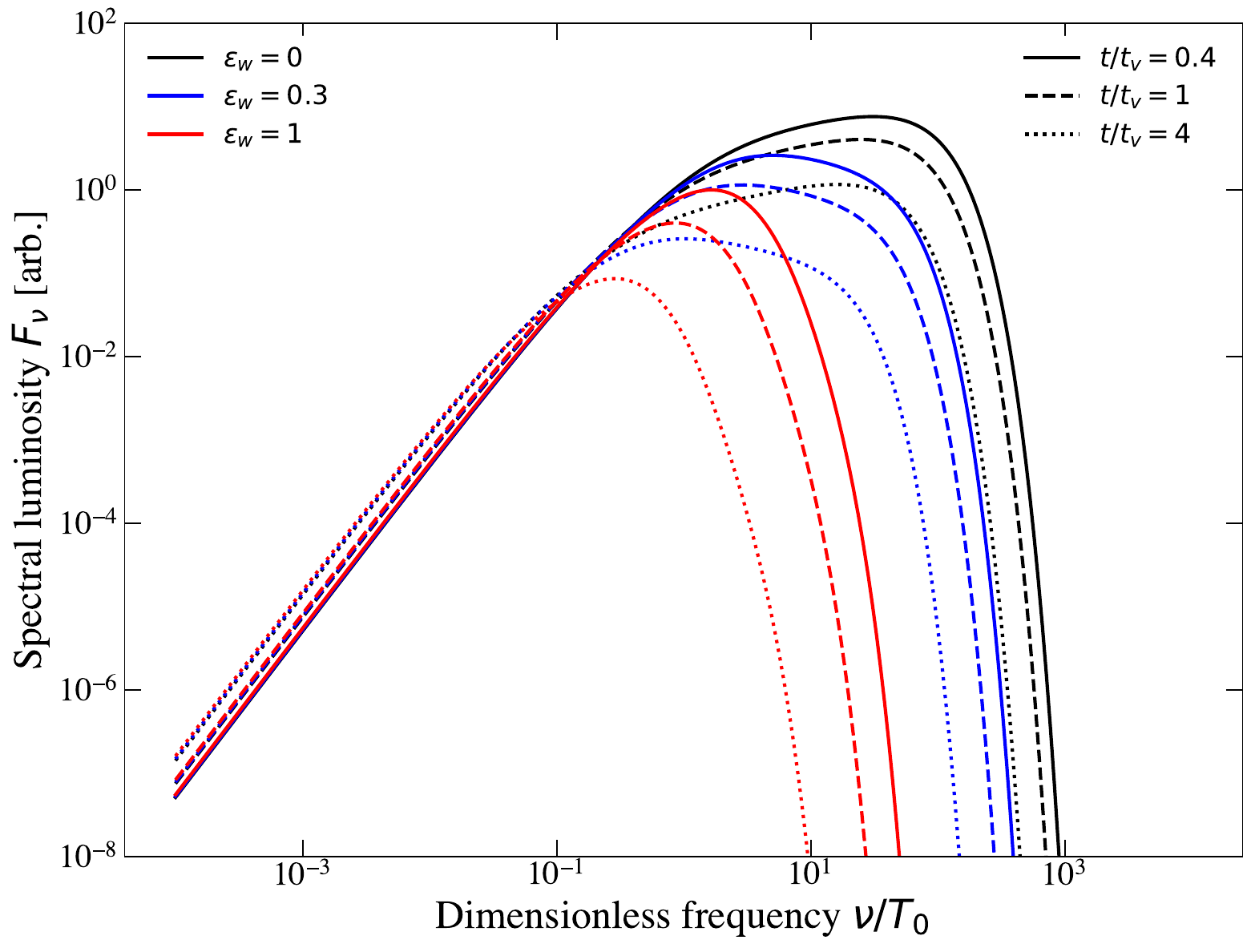}
    \caption{{\bf Left:} the evolving bolometric luminosity (arbitrarily but consistently normalised) of Green's function solutions of the coupled inflow-outflow disk equation for different values of the outflow parameter $\varepsilon_w$ (see legend), plotted as a function of dimensionless time. {\bf Right:} the (arbitrarily but consistently) normalised spectral energy distributions of the Green's function solutions plotted in Figure \ref{fig:dens+temp}, plotted against dimensionless observing frequency. Outflow parameters $\varepsilon_w$ are denoted by color, while the dimensionless times are denoted by line style. Outflows strongly suppress high energy emission, resulting from the flattened disk temperature profiles. Outflows can also modify the mid frequency spectral slope of the disk, but do not strongly impact emission in the Rayleigh Jeans tail of the disk.    }
    \label{fig:Lbol+Spec}
\end{figure*}
For these global wind efficiency solutions, the rate of decay of the bolometric luminosity is enhanced relative to their windless ($\varepsilon_w=0$) cousins, with a late time decay index (see Figure \ref{fig:Lbol+Spec})
\begin{equation}
    \log L_{\rm bol} \propto -\left({4-2\mu + 4\varepsilon_w\over 3 - 2\mu}\right) \log(t),
\end{equation}
for stress parameterization $W^r_\phi\propto r^\mu$. The self-similar turbulent stress solution $\mu=0$ is a particularly natural example, and we see that the decay in the intrinsic luminosity of the disk is enhanced from $4/3$ to $4(1+\varepsilon_w)/3$ by the presence of outflows. 

\begin{figure*}
    \centering
    \includegraphics[width=0.49\linewidth]{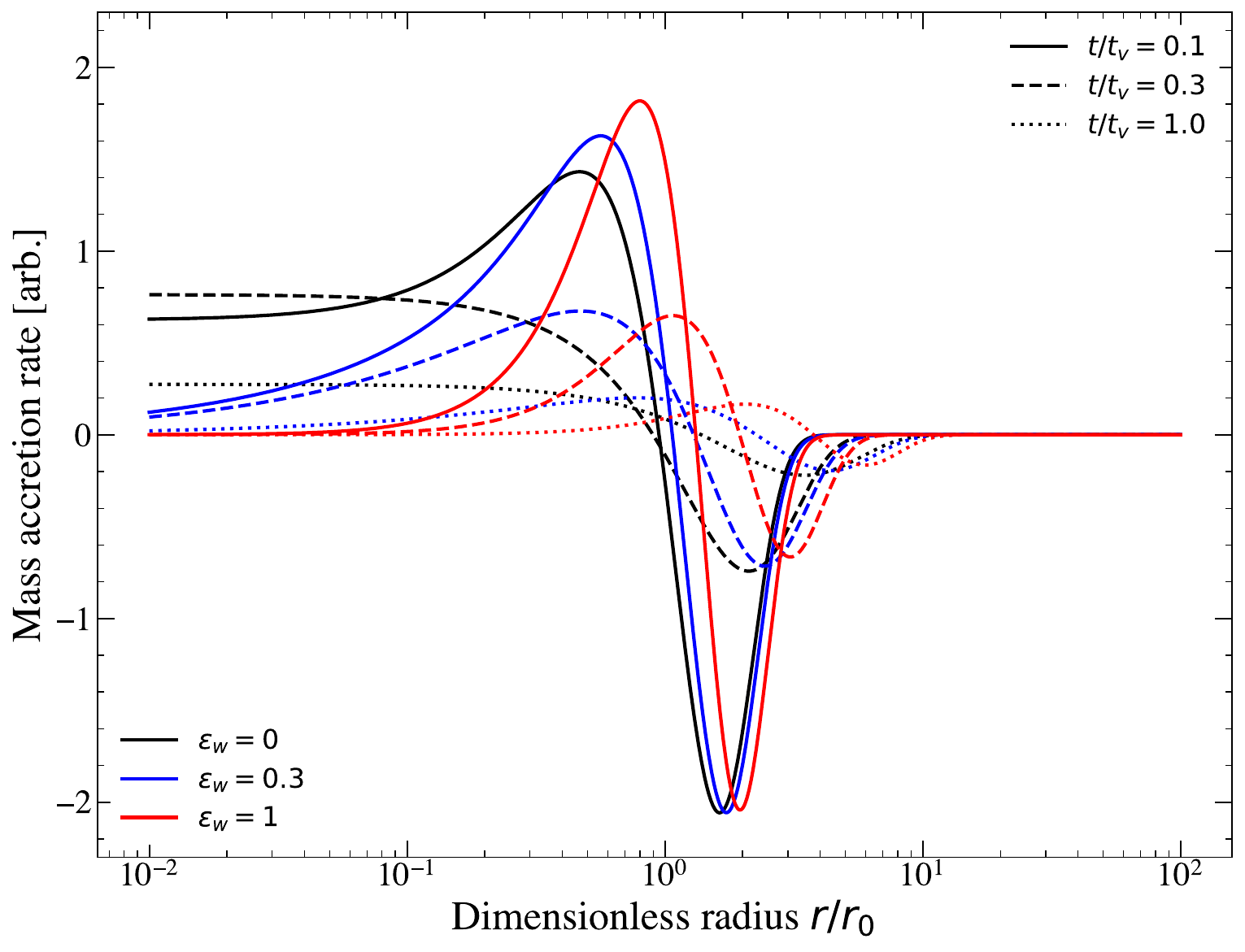}
    \includegraphics[width=0.49\linewidth]{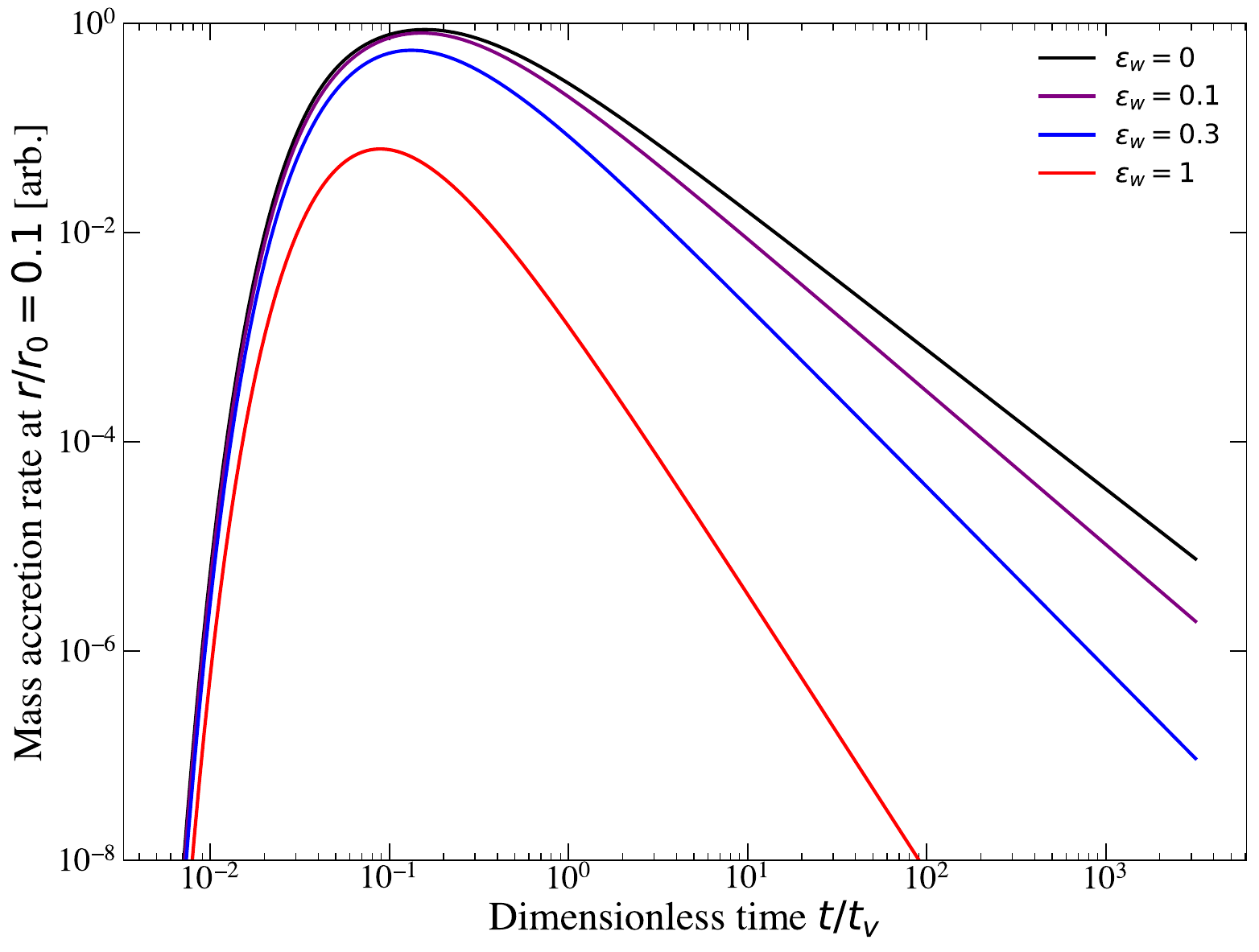}
    \caption{The accretion rate (arbitrarily but consistently normalised) resulting from Green's function solutions of the coupled inflow-outflow disk equations for different values of the wind efficiency $\varepsilon_w$. {\bf Left:} the spatial dependence of the accretion rate for different times (denoted by line styles) and efficiencies (denoted by color). Note the strong suppression of the inner disk accretion rate for large wind efficiencies. {\bf Right:} the evolution (in dimensionless time) of the accretion rate at $r/r_0=1/10$, showing both the suppression as well as the more rapid time evolution for larger wind efficiencies. }
    \label{fig:Mdot}
\end{figure*}

\begin{figure*}
    \centering
    \includegraphics[width=0.49\linewidth]{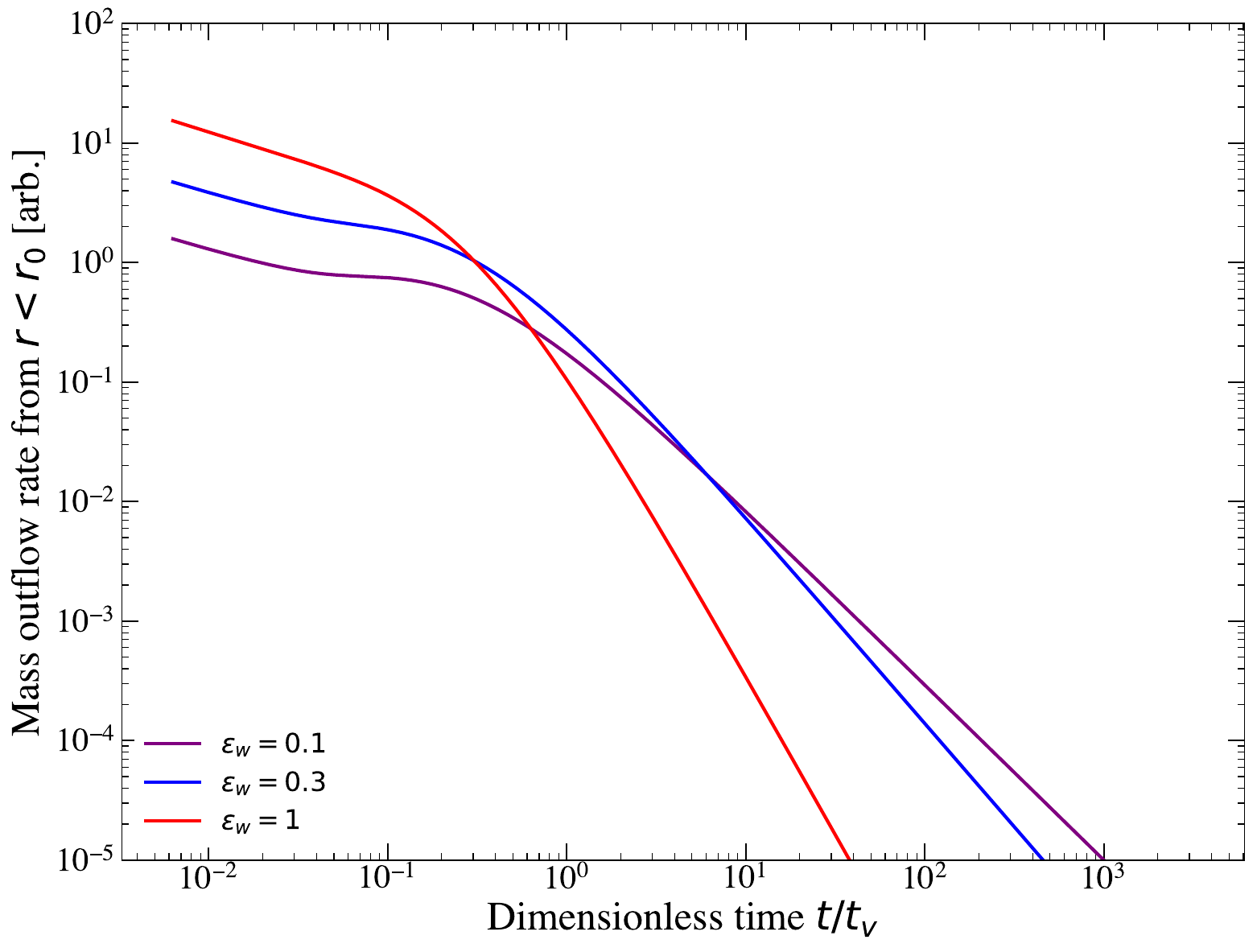}
    \includegraphics[width=0.49\linewidth]{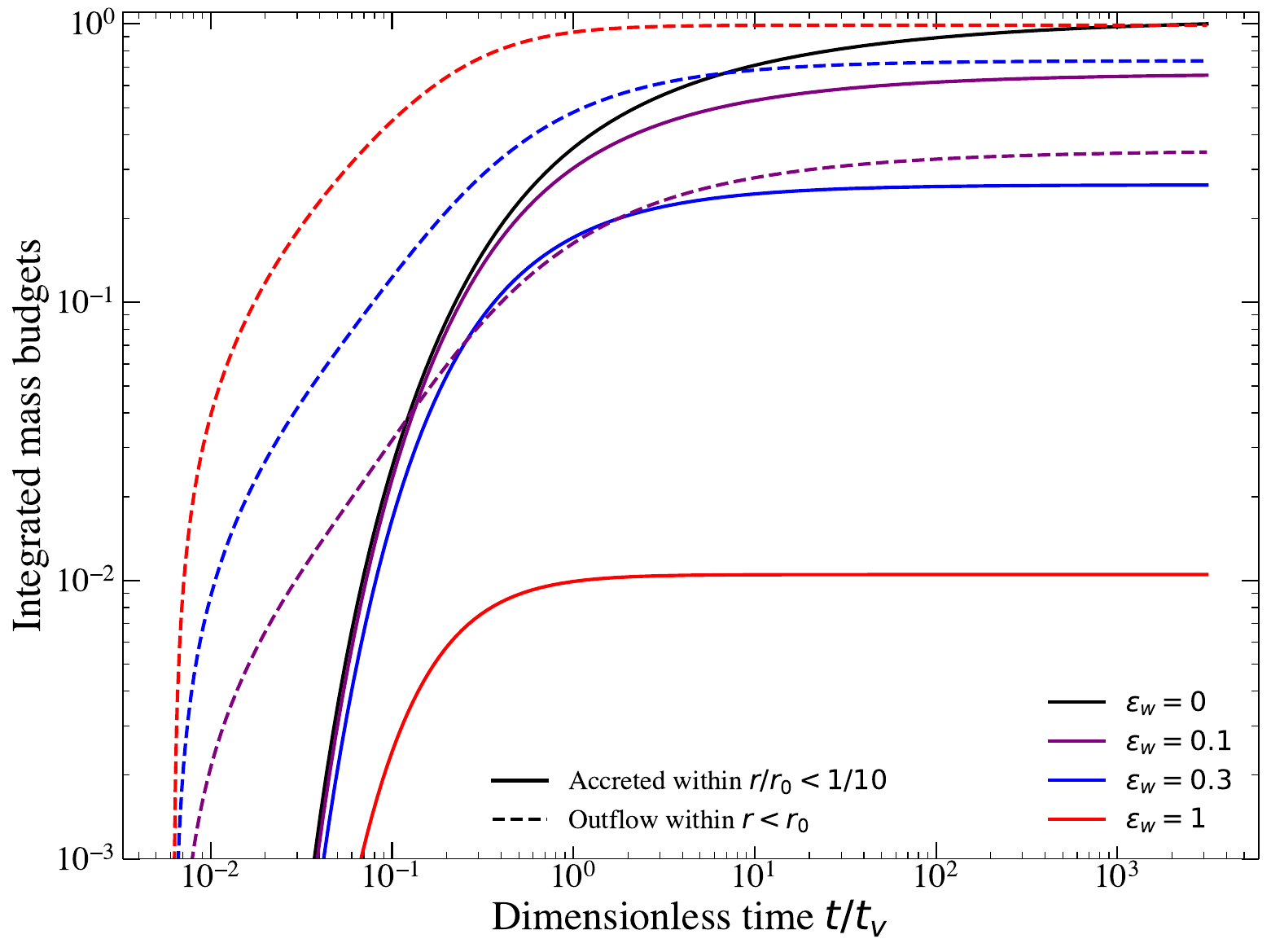}
    \caption{{\bf Left:} the mass outflow rate from all radius interior to the initial radius of the flow (arbitrarily but consistently normalised) for different wind efficiencies (denoted by color), as a function of dimensionless time. Mass outflow peaks at early times, when the accretion rate is highest. {\bf Right:} the integrated mass budgets for inflow (matter which passes through $r = r_0/10$, solid curves) and outflow (matter launched from $r < r_0$, dashed curves) as a function of dimensionless time. The curves are normalised to sum to the accreted mass in the absence of winds. We see that even moderate wind efficiencies can prevent a large fraction of matter from reaching small radii.    }
    \label{fig:Mout}
\end{figure*}

This bolometric luminosity is still temporally coupled to the accretion rate in the inner disk, which scales like
\begin{multline}
    \log \dot M_{\rm acc}(r, t) \propto 2\varepsilon_w \log(r)\\ -\left({4-2\mu + 4\varepsilon_w\over 3 - 2\mu}\right) \log(t),
\end{multline}
although we note that the accretion rate onto the origin is formally zero for any non-zero outflow (see Figure \ref{fig:Mdot}). Naturally, the mass outflow rate from all radii smaller than $r$ shows the same scaling with radius and time as the local accretion rate at $r$ (a result of mass conservation in the asymptotic large time limit, Figure \ref{fig:Mout}). 

This enhanced luminosity decay can be readily understood on simple physical grounds: the dissipation in the disk (and therefore the source of the disk luminosity) is still dominated by the innermost regions, but the amount of matter which reaches down to these regions is strongly suppressed owing to all of the matter lost in outflows. 

\subsection{Disk expansion}
The outer radius of the disks in these Green's function solutions grow with time, and in fact follow an identical expansion rate to that seen in the absence of winds, namely 
\begin{equation}
    \log R_{\rm out}(t) \propto \left({2 \over 3 - 2\mu}\right) \log (t). 
\end{equation}
Again, this is the general result for $W^r_\phi \propto r^\mu$, which can be derived from noting the typical radial location (as a function of time) at which the exponential cut off in the Green's functions begins to dominate the behavior of the solution. This is a potential surprising result, as it even holds in the limit in which the winds apply non-zero torques onto the flow (see Figure \ref{fig:dens+temp}). 

What is different from the wind-less limit is that the expansion of the disk is no longer directly coupled to the accretion rate onto the central object. Indeed, for no winds, one has $\dot M_{\rm acc} \sim t^{-n} \to R_{\rm out} \sim t^{2n-2}$. When outflows are present, this one to one coupling between accretion rate and expansion rate is broken. 

The expansion of compact accretion flows has important observational implications for those disks formed in tidal disruption events. It is important to note that the presence of outflows does not modify the ``viscous'' expansion of these flows.

\subsection{Disk temperature profiles}
The temperature profile of the disks with constant wind efficiencies are flattened, where the temperature profile follows (Figure \ref{fig:dens+temp})
\begin{equation}
    \log(T) \propto -\left({3\over 4} - {1 \over 2}\varepsilon_w \right) \log r . 
\end{equation}
Heuristically, this can be understood from the quasi-steady state argument (this is of course not rigorous, but highlights the scalings) 
\begin{equation}
   \sigma T^4 \sim {GM\dot M_{\rm acc} \over 8\pi r^3}, 
\end{equation}
with the result derived above for the scaling of the accretion rate with radius $\dot M_{\rm acc} \sim r^{2\varepsilon_w}$. This will have various observational implications. Firstly, emission usually sourced in the innermost disk regions (e.g., thermal X-ray emission) will be strongly suppressed by a moderate $\varepsilon_w$, as the innermost disk regions will no longer reach large enough temperatures to produce this thermal flux (note that this X-ray suppression argument is independent of any absorption of high energy photons in the outflow itself, which could further suppress the observed emission; see Figure \ref{fig:Lbol+Spec}). 

Another implication of this flattened temperature profile would be a change in the mid-frequency spectral slope of these disks. The intrinsic spectrum of an accretion flow (i.e., neglecting any impacts from absorption and reprocessing of photons, etc., on their path to the observer) is given by 
\begin{equation}
    F_\nu(\nu, t) = \nu^3\int_0^\infty {r\, {\rm d}r \over \exp(\nu/T(r, t)) -1}, 
\end{equation}
where we have set various fundamental constants to $1$. Figure \ref{fig:Lbol+Spec} highlights the important impacts that accretion disk outflows can have on both the frequency-dependence and amplitude of this intrinsic spectrum.  

It is an elementary calculation to show that if every disk radius emits as a pure blackbody (i.e., the above, while this is an oversimplification it highlights the general point), with temperature profile $T\propto r^{-q}$, then the composite spectrum observed at frequencies between $kT_{\rm min}\ll h\nu \ll kT_{\rm max}$ has slope 
\begin{equation}
    \log F_\nu \propto \left(3 - {2\over q}\right) \log \nu, 
\end{equation}
or in this instance 
\begin{equation}
    \log F_\nu \propto \left({1-6\varepsilon_w\over3 - 2\varepsilon_w}\right) \log \nu . 
\end{equation}
Note that of course, for high outflow efficiencies (and therefore very shallow temperature profiles), the frequency range which actually satisfies $kT_{\rm min}\ll h\nu \ll kT_{\rm max}$ can be very narrow and indeed an obvious ``mid frequency'' part of the spectrum may not even exist (e.g., the $\varepsilon_w = 1$ solution in Figure \ref{fig:Lbol+Spec}).  

Observations of the mid-frequency regions of accretion flows could in principle be able to place constraints on outflow efficiencies. 

\subsection{Variability in the accretion rate}
\begin{figure*}
    \centering
    \includegraphics[width=0.49\linewidth]{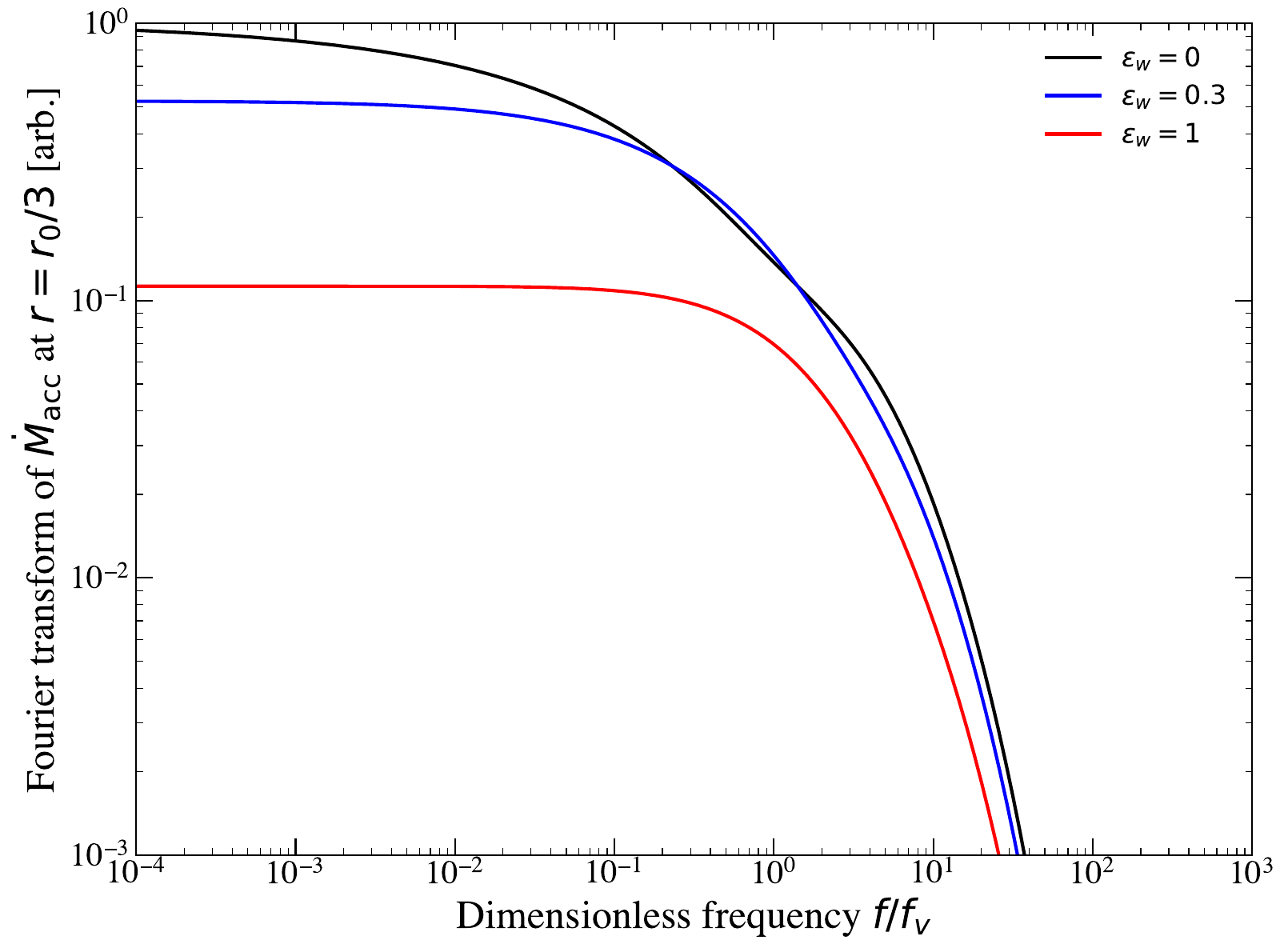}
    \includegraphics[width=0.49\linewidth]{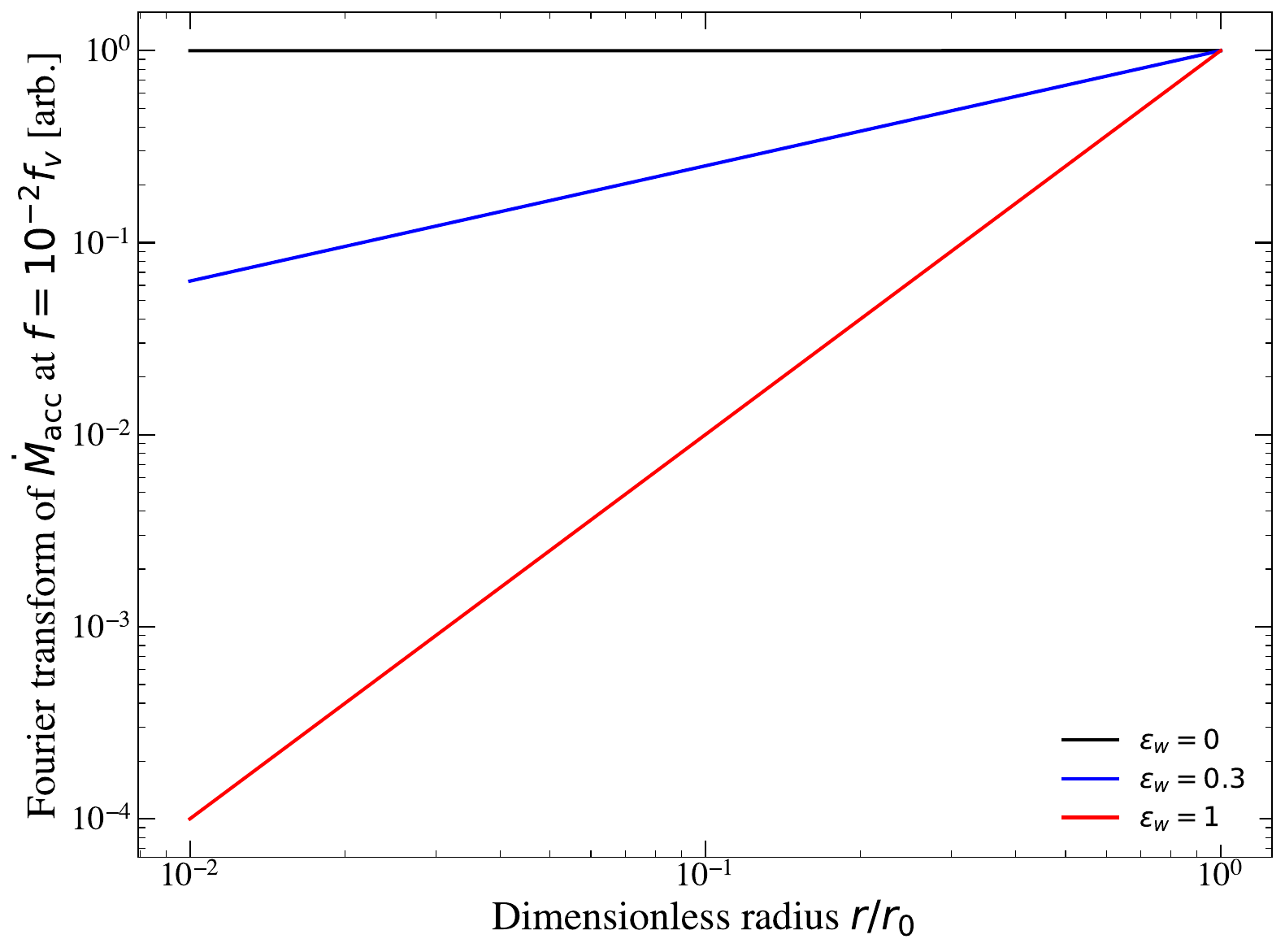}
    \caption{The Fourier transform of the mass accretion rate for different wind efficiencies (denoted by color). {\bf Left:} the (absolute value of) the variability in the mass accretion rate at $r/r_0=1/3$, showing that outflows generically suppress variability in the accretion rate. {\bf Right:} the radial dependence of the accretion rate variability for different wind efficiencies, at low frequencies compared to the local viscous frequency $f = 0.01/t_v$. Showing that the larger the separation between two disk radii the stronger the suppression in the accretion rate variability (on long timescales) in the presence of outflows. }
    \label{fig:fourier}
\end{figure*}
Finally, we constructed the Fourier transform of the Green's functions in the presence of outflows, highlighting how outflows impact the variability properties of accretion disks. 

Our two main variability results relevant for observations are the following, firstly outflows suppress the variability in the mass accretion rate at small radii 
\begin{equation}
    \log \widetilde {\dot M}_{\rm acc} (r\ll r_0, f) \propto {2\varepsilon_w} \log r, 
\end{equation}
where $\widetilde{\dot M}_{\rm acc}$ is the Fourier transform of the accretion rate (Figure \ref{fig:fourier}). Variability of the accretion rate is also suppressed at low frequencies (long timescales), by 
\begin{equation}
    \log \widetilde {\dot M}_{\rm acc} (r < r_0, f\to 0) \propto {2\varepsilon_w} \log r - \varepsilon_w \log(r_0). 
\end{equation}
Both of these results can be readily understood physically, and result simply from the outflows (i) reducing ``communication'' between inner and outer disk regions (as material is removed from the system on its way to the center), and (ii) suppressing the accretion rate generally. At high Fourier frequencies the disk variability is still suppressed by an exponential factor $\exp(-Af^{1/2})$, a result of the diffusive nature of disk turbulence. 

One may speculate that this suppression in the inner disk accretion rate variability may lead to super-Eddington accretion flows (which likely have larger $\varepsilon_w$) producing less variable emission. 

\section{Conclusions}

The purpose of this paper has been to construct explicit and simple Green's function solutions to the Newtonian thin disk evolution equation in the presence of outflows (sink terms for matter) which may or may not impart a torque on the disk (sink or source terms for angular momentum), and to examine the physical implications of outflows on the local and global disk structure. 

Ultimately, we are interested in constructing solutions to a time dependent relativistic theory of black hole accretion in the presence of outflows. We have gone into such detail of the properties of Newtonian disk solutions because, while relativistic solutions are required for a detailed {\it quantitative} comparison to observations of accreting black holes, the {\it qualitative} nature of such solutions inherit many properties of their Newtonian counterparts (particularly in the outer disk). Therefore many of the observational consequences of outflows (discussed in detail in section 7) will directly carry over to the relativistic regime in a qualitative sense. Of particular observational interest will be (i) the more rapid decay of the bolometric luminosity of the accretion flow in the presence of outflows; (ii) the flattening of the disk temperature profile and its associated modifications to the mid-frequency spectral slope (making the disk spectrum redder), and the high-energy disk spectrum (suppressing high energy emission); and (iii) the suppression of accretion rate variability at small radii and low frequencies. 

On a technical level, knowing explicit solutions to the Newtonian disk equation is an essential step in constructing asymptotic solutions of the more algebraically complicated relativistic disk equations. This will be the subject of the next paper in this series. 

It seems likely that an explicit knowledge of Green's function solutions of the Newtonian disk equations in the presence of outflows will be of interest to the proto-planetary disk community.

\section*{Note}
In the final week of preparation of this manuscript, M. Tamilan presented similar calculations which overlap with some of the solutions presented in this work (a subset of the surface density Green's functions for radius-dependent wind timescales, see \citealt{Tamilan25}). We have considered more general wind launching profiles in this work, while Tamilan considered more general boundary conditions. I am grateful to J. Matthews for bringing this paper to my attention. 

\section*{Acknowledgments}
A. Mummery is supported by  by the John N. Bahcall Fellowship Fund at the Institute for Advanced Study. 

\bibliographystyle{aasjournal}
\bibliography{andy}

\appendix
\section{Normalised Green's functions}
Every Green's function for the combination $\xi \equiv r \Sigma W^r_\phi$ in this paper has the same general form, namely 
\begin{equation}
    G(r, r_0, t) = A {(r/r_0)^{{1\over 4} + a} \over (t/t_{v, 0})} \exp\left({-1 - (r/r_0)^{2b} \over 4 (t/t_{v, 0})}\right) I_c\left({(r/r_0)^b \over 2(t/t_{v, 0})}\right) .
\end{equation}
Where $a, b$ and $c$ differ depending on one's choice of wind parameterisation. Taking the $t\to 0$ limit of this expression, and using the large $z$ expansion of the Bessel function $\lim_{z\to\infty}I_\nu(z) \to e^z/\sqrt{2\pi z}$, one finds 
\begin{equation}
    G(r, r_0, t\to0) \to 2 A (r/r_0)^{{1\over 4} + a - {b\over 2}} \times \lim_{t\to0} \left\{{1\over \sqrt{2\pi (2t/t_{v, 0})}} \exp\left(-{(1-(r/r_0)^b)^2\over 2(2t/t_{v, 0})}\right)\right\} .
\end{equation}
We recognize the limiting delta function definition in the final term, leading to 
\begin{equation}
    G(r, r_0, t\to0) \to {2 A (r/r_0)^{{1\over 4} + a - {b\over 2}} \over b} \times \delta(1-(r/r_0)).
\end{equation}
Integrating this over the whole disk, where $W^r_\phi \equiv w (r/r_0)^\mu$, one finds 
\begin{equation}
    \int_0^\infty G(r, r_0, t\to0)\, {\rm d}r = {2Ar_0\over b} = \int_0^\infty r \Sigma(r, r_0, t\to0)\, w (r/r_0)^\mu {\rm d}r = {w M_d \over 2\pi},
\end{equation}
where $M_d$ is the initial disk mass. Therefore $A = wb M_d / 4\pi r_0$. The mass accretion rate Green's function is given by 
\begin{equation}
    G_{\dot M_{\rm acc}} (r, r_0, t) =  2\pi r \Sigma U^r = - 2\pi \sqrt{4r\over GM} {\partial \over \partial r} G(r, r_0, t) - 2\pi \sqrt{4r^3\over GM}F_J = - {4\pi \over \sqrt{GMr_0}} x^{1/2} {\partial \over \partial x} G(x, t)  - 2\pi \sqrt{4r^3\over GM}F_J, 
\end{equation}
where $x\equiv r/r_0$, and $F_J$ depends on the specific problem at hand (but can be constructed from the Green's function for the surface density). Clearly we require the derivative 
\begin{equation}
    Y = {x^{1/2} \over \tau} {\partial \over \partial x} \left(x^{{1\over 4}+ a} \exp\left({-1-x^{2b} \over 4\tau}\right) I_c\left({x^b\over 2\tau}\right)\right),
\end{equation}
where $\tau \equiv t/t_{v, 0}$. This is equal to 
\begin{equation}
    Y = {x^{a-1/4}\over \tau} \left[{b x^{b}   \over 2\tau}I_{c - 1}\left({x^b\over 2 \tau}\right) + \left(a +{1\over 4} - bc - {bx^{2b}\over 2\tau}\right)  I_{c}\left({x^b\over 2 \tau}\right)\right]\exp\left({-1-x^{2b} \over 4\tau}\right).
\end{equation}
And so the mass accretion rate Green's function is 
\begin{equation}
    G_{\dot M_{\rm acc}}(r, r_0, t) = - {4\pi A \over \sqrt{GMr_0}} Y(x, \tau) - 2\pi \sqrt{4r^3\over GM} F_J. 
\end{equation}
\newpage
\section{Different stress profiles}
We now consider the case $W^r_\phi = w(r/r_0)^\mu$. In this limit one requires $t_w=t_{w, 0}(r/r_0)^{3/2-\mu}$ for the existence of simple analytical solutions in the parameterized wind time case. The below parameters are to be substituted into the form of the general profile introduced above. In each case the bolometric luminosity decays with index $n = 1+c$ (i.e., $L \sim t^{-n}$), while the temperature profile has index $m = 13/16 - (a + bc)/4$, i.e., $T \sim r^{-m}$. 
\subsection{Zero torque winds}
In this limit we have 
\begin{align}
    t_{v, 0} &= {2\over (3-2\mu)^2} {\sqrt{GMr_0^3} \over w}\\
    \tilde t & = {w t_{w, 0}\over 2\sqrt{GMr_0^3}} = {1\over (3-2\mu)^2} {t_{w, 0}\over t_{v, 0}}, \\
    a &= 0, \\
    b &= {3-2\mu \over 4}, \\
    c &= {1\over 3-2\mu} \sqrt{1 + {8\over \tilde t}}. 
\end{align}
\subsection{Torquing winds}
For $\Delta U_\phi = \Gamma U_\phi$ we have
\begin{align}
    t_{v, 0} &= {2\over (3-2\mu)^2} {\sqrt{GMr_0^3} \over w}\\
    \tilde t & = {w t_{w, 0}\over 2\sqrt{GMr_0^3}} = {1\over (3-2\mu)^2} {t_{w, 0}\over t_{v, 0}}, \\
    a &= -{\Gamma \over 2\tilde t}, \\
    b &= {3-2\mu \over 4}, \\
    c &= {1\over 3-2\mu} \sqrt{1 + {4\Gamma^2 + 4 \Gamma \tilde t+ 8\tilde t \over \tilde t^2}}. 
\end{align}
\subsection{Constant wind efficiency}
For a global constant $\varepsilon_w$ we have 
\begin{align}
    t_{v, 0} &= {2\over (3-2\mu)^2} {\sqrt{GMr_0^3} \over w}\\
    a &= \varepsilon_w, \\
    b &= {3-2\mu \over 4}, \\
    c &= {1 + 4 \varepsilon_w \over 3-2\mu} . 
\end{align}
\end{document}